\documentclass[aps,prl,reprint,longbibliography]{revtex4-1}

\usepackage{amsmath}
\usepackage[separate-uncertainty=true]{siunitx}
\usepackage{graphicx}
\graphicspath{{figures/}}

\renewcommand{\vec}[1]{\boldsymbol{#1}}
\newcommand{\abs}[1]{\left\vert #1 \right\vert}

\begin{document}

\title{Above and beyond: Holographic tracking of 
axial displacements in holographic optical tweezers}

\author{Michael J. O'Brien}
\author{David G. Grier}

\affiliation{Department of Physics and Center for Soft Matter Research,
  New York University, New York, New York 10003, USA}

\begin{abstract}
How far a particle moves along the optical axis in a holographic
optical trap is not
simply dictated by the programmed motion of the trap,
but rather depends on an interplay of the trap's changing
shape and the particle's material properties.
For the particular case of colloidal spheres in 
optical tweezers,
holographic video microscopy reveals that
trapped particles tend to move farther along the axial 
direction than the traps that are moving them
and that different kinds 
of particles move by different amounts. 
These surprising and sizeable variations in
axial placement can be explained by 
a dipole-order theory for optical forces.
Their discovery highlights the need for
real-time feedback to achieve precise control
of colloidal assemblies in three dimensions
and demonstrates that holographic microscopy
can meet that need.
\end{abstract}

\maketitle

\section{Introduction}
\label{sec:introduction}

Holographic optical traps use the forces and torques
exerted by computer-generated holograms to
localize and manipulate micrometer-scale objects
\cite{dufresne98,grier03}.
In principle, holographic traps can move
colloidal particles along arbitrary 
paths in three dimensions
and can arrange multiple particles into
precisely specified three-dimensional configurations
\cite{roichman05,chapin06,benito2008constructing}.
In practice, however, where a trap places a particle
depends on details of the particle's interaction with the light field.
Here, we use Lorenz-Mie microscopy to measure colloidal spheres'
trajectories in holographic traps and thereby
to demonstrate that particles with different sizes and compositions
not only reside at different axial positions within
coplanar traps, but indeed travel substantially
different distances
when the traps are displaced along the optical axis.
This surprising observation can be explained by considering how 
an optical trap's structure depends on its axial position.
Even so, axial displacements pose a practical
challenge because variations from particle to
particle can be large and are difficult to predict
quantitatively.
We demonstrate that Lorenz-Mie microscopy
can provide the real-time feedback needed
to achieve precise three-dimensional control
over colloidal assemblies with holographic optical
traps.

\section{Holographic optical trapping}
\label{sec:hot}

\begin{figure*}[t]
    \centering
    \includegraphics[width=\textwidth]{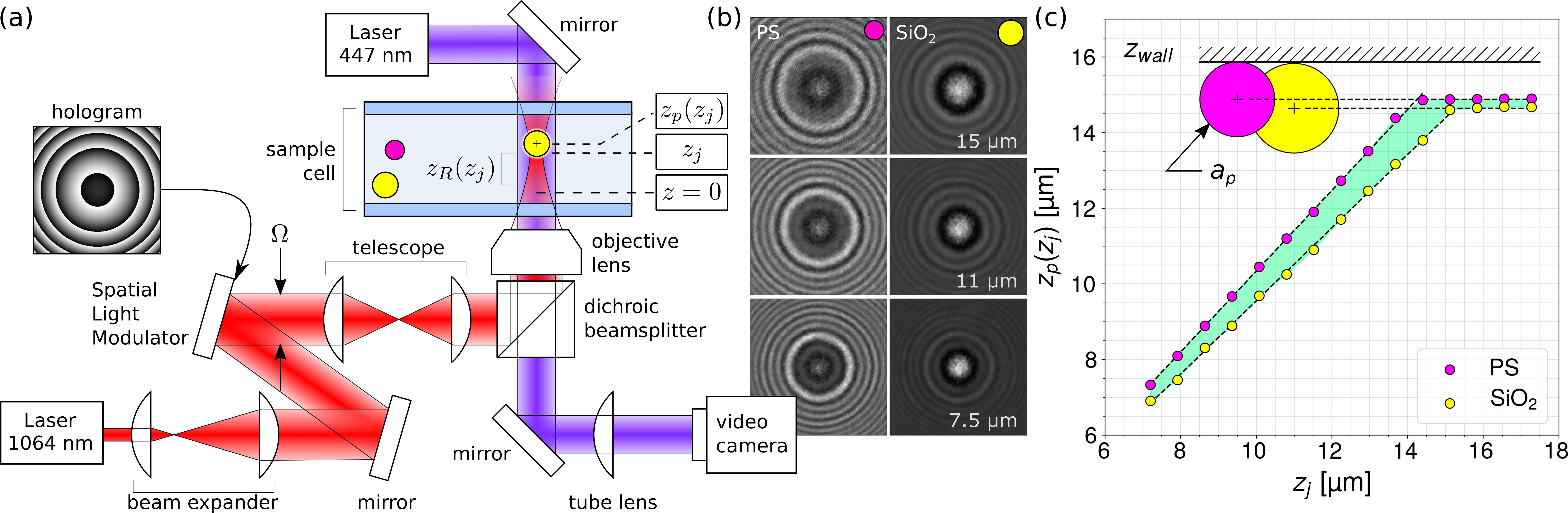}
    \caption{(a) Schematic representation of the combined instrument
      for holographic optical trapping and in-line holographic video
      microscopy.
      Holographic traps are projected into
      the sample cell by imprinting a phase hologram onto the wavefronts of an
      infrared laser beam and relaying the hologram to
      the input pupil of an objective lens with a dichroic
      beamsplitter. In-line holograms of trapped particles
      are recorded with a blue laser beam that passes
      through the dichroic to a video camera.
      (b) Typical holograms of a polystyrene sphere (PS) and a silica
      sphere
      (SiO$_2$) displaced to specified
      axial positions, $z_j$, by adjusting the phase hologram.
      (c) Measured axial positions, $z_p(z_j)$, 
      of a polystyrene sphere (PS)
      and a silica sphere (SiO$_2$)
      as a function of specified trap
      position, $z_j$. Large circles 
      depict the holographically measured
      radii, $a_p$, of the two spheres and are positioned
      at the measured plateau heights of
      their trajectories. The two particles
      thus agree on the height, $z_\text{wall}$,
      of the upper glass wall of their sample cell.
      Shading between
      the traces emphasizes the spheres'
      increasing axial separation.}
    \label{fig:hothvm}
\end{figure*}

The holographic trapping technique,
depicted schematically in Fig.~\ref{fig:hothvm}(a),
uses computer-generated
holograms to structure the wavefronts of a laser beam
so that the modified beam forms the desired configuration of optical
traps when brought to a focus by a strongly converging
objective lens \cite{dufresne98,grier03}.
The implementation used for this study is driven by
a \SI{10}{\watt} fiber laser (IPG Photonics, YLR-10-LP)
operating at a vacuum wavelength of 
$\lambda_0 = \SI{1064}{\nm}$.
Holograms are imprinted on this laser's wavefronts
using a liquid-crystal spatial light modulator
(Holoeye Pluto).
The modified beam is then projected into the sample
by a $100\times$
oil-immersion objective lens
with a focal length of $f = \SI{200}{\um}$ and a
numerical aperture of $\mathrm{NA} = 1.4$.
(Nikon S-Plan Apo).

The ideal scalar hologram encoding
$N$ point-like optical tweezers \cite{curtis02,polin05},
\begin{equation}
    E(\vec{\rho})
    =
    \sum_{j = 1}^N
    E_j
    \exp\left(-i \frac{k}{f} \, \vec{r}_j \cdot \vec{\rho}\right)
    \exp\left( i \frac{k}{2 f^2} z_j \rho^2 \right),
    \label{eq:superposition}
\end{equation}
places the $j$-th trap at position
$\vec{r}_j$ relative to the center of the objective's
focal plane at $z = 0$.
Here, $\vec{\rho}$ is the two-dimensional
coordinate of a point
in the hologram plane, 
$k = 2 \pi n_m / \lambda_0$ is the light's wavenumber in
a medium of refractive index $n_m$, and $E_j$ is the
complex amplitude of the $j$-th trap.
The total power required to project this
pattern is
\begin{equation}
    \label{eq:power}
    P
    =
    \frac{1}{2} \Omega \, n_m c \epsilon_0 \, \sum_{j=1}^N \abs{E_j}^2,
\end{equation}
where
$c$ is the speed of light in vacuum,
$\epsilon_0$ is the vacuum permittivity,
and $\Omega$
is the effective cross-sectional area of the
projection system.
The example hologram in Fig.~\ref{fig:hothvm}(a)
encodes the phase profile,
\begin{equation}
    \varphi(\vec{\rho})
    =
    \frac{k}{2 f^2} \, z_j \, \rho^2 \bmod 2 \pi ,
\end{equation}
of a single optical tweezer displaced
by $z_j$ along the
optical axis.

Equation~\eqref{eq:superposition} can be generalized to
project line traps \cite{roichman06,roichman06c},
ring traps \cite{roichman07b,shanblatt11}, 
knotted traps \cite{shanblatt11,rodrigo13} and
tractor beams \cite{lee10,ruffner12a}, 
among many other modalities.
Hardware-accelerated algorithms 
can perform the necessary field
calculations in real time, permitting dynamic interaction with
trapped materials \cite{bianchi2010real}.

Although the focal point of a trap may be located at $\vec{r}_j$,
the point of mechanical equilibrium for a
particle localized in the trap typically
is displaced by radiation pressure \cite{ashkin86}
as well as by external forces such as gravity.
Measuring this effect with
Lorenz-Mie microscopy \cite{lee07a} not only reveals material-dependent
displacements, but also shows that the displacement
depends strongly on the trap's axial position, which means
that the particle systematically moves either more or less than the trap
that is moving it.

\section{Lorenz-Mie microscopy}
\label{sec:hvm}

The instrument depicted in Fig.~\ref{fig:hothvm}(a)
creates in-line holograms
of optically trapped
particles \cite{sheng06,lee07,lee07a}
by illuminating the sample with
a collimated laser beam at vacuum wavelength
$\lambda_1 = \SI{447}{\nm}$ (Coherent Cube)
aligned with the optical axis of the objective lens.
Light scattered by a particle interferes with the
rest of the beam in the focal plane of the objective lens
and is relayed by a tube lens to a monochrome video camera
(Flir Flea3) that records
the intensity of the magnified interference pattern.
The imaging subsystem is separated from
the trapping subsystem by a dichroic
beamsplitter (Semrock).
The images in Fig.~\ref{fig:hothvm}(b) show
holograms of a polystyrene sphere and a silica sphere 
recorded at three different axial displacements, $z_j$.

Lorenz-Mie microscopy is distinguished from other
holographic microscopy techniques by the approach used
to extract information from recorded holograms.
The incident field is modeled as a 
monochromatic plane wave linearly polarized along $\hat{x}$ and
propagating along $-\hat{z}$:
\begin{equation}
    \vec{E}_0(\vec{r}, t) = E_0 e^{-i q z} e^{-i \omega t} \hat{x},
\end{equation}
where $q = 2 \pi n_m/\lambda_1$ is the wavenumber
of the imaging illumination and $\omega = 2\pi c/\lambda_1$ is its frequency.
A particle located at position $\vec{r}_p$ within this beam
creates a scattered field,
\begin{equation}
    \vec{E}_s(\vec{r}, t) 
    =
    E_0 \, e^{-i q z_p} \,
    \vec{f}_s(q (\vec{r} - \vec{r}_p)),
\end{equation}
where $\vec{f}_s(q\vec{r})$ is the Lorenz-Mie scattering function
\cite{bohren83,mishchenko02}.
If the particle is sufficiently small, the intensity, $I(\vec{r})$, recorded
at point $\vec{r}$ in the camera plane may be modeled as
\cite{lee07a}
\begin{equation}
    b(\vec{r})
    = \frac{I(\vec{r})}{\abs{E_0}^2}
    =
    \abs{\hat{x} + 
    e^{-i q z_p} \, \vec{f}_s(q ( \vec{r} - \vec{r}_p))}^2 .
    \label{eq:br}
\end{equation}

For spherical scatterers, the Lorenz-Mie
function is parameterized by the sphere's
radius, $a_p$, and its refractive index, $n_p$.
Fitting Eq.~\eqref{eq:br} pixel-by-pixel to a recorded
hologram therefore measures a particle's three-dimensional
position and its size while also providing insight into its
composition through the refractive index.
Published realizations of this technique demonstrate
nanometer precision for in-plane tracking,
five-nanometer precision for axial tracking,
part-per-thousand precision for $n_p$ and 
five-nanometer precision for $a_p$ \cite{lee07a,krishnatreya14}.
Lorenz-Mie microscopy, moreover, can track a particle
as it moves over large axial ranges without requiring
mechanical scanning.
Comparably good all-optical axial tracking has been
achieved with dynamically focused stereomicroscopy \cite{lee2014dynamic},
which uses a spatial light modulator
to maintain a particle in optimal focus.
The present study relies on the ability of
Lorenz-Mie microscopy to track particles
at different axial positions simultaneously
while also measuring their radii and
identifying their compositions through their
refractive indexes.

The only instrumental calibration constants for
Lorenz-Mie microscopy are the vacuum wavelength
of the laser, the magnification of the microscope and
the refractive index of the medium.
Hardware-accelerated computation of the Lorenz-Mie scattering
function enables complete fits to be performed in tens of
milliseconds \cite{cheong09,yevick14,hannel18}, 
which is fast enough for interactive operation
and instrumental feedback.

\section{Above and beyond: Axial displacements}
\label{sec:abovebeyond}

\begin{figure*}
    \centering
    \includegraphics[width=\textwidth]{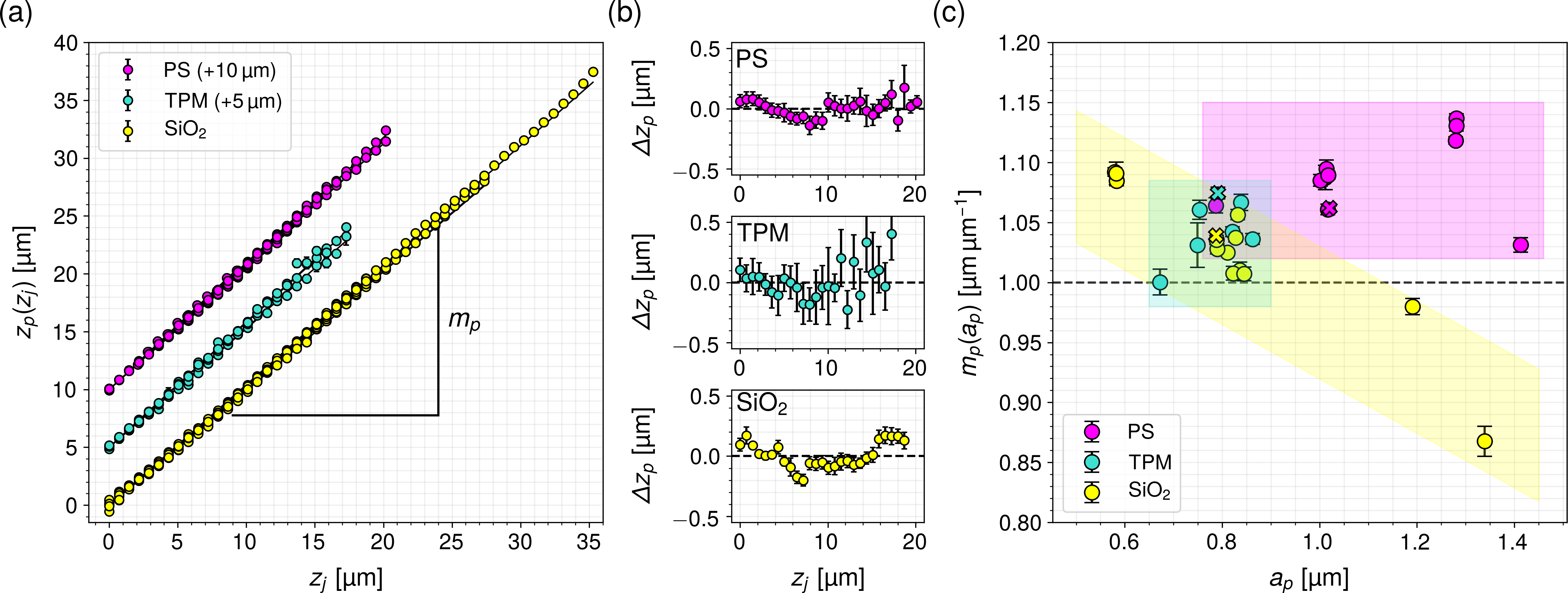}
    \caption{(a) Axial displacements of colloidal
    spheres in holographic optical traps, including
    polystyrene (PS), 3-(trimethoxysilyl)propyl methacrylate (TPM) and silica (SiO$_2$).
    Data for polystyrene and TPM are displaced upward
    for clarity. Each composition is represented by data from
    multiple spheres (11 for PS, 10 for SiO$_2$ and 4 for TPM)
    of nominally identical radii
    measured in different sample cells.
    (b) Residuals of displacements
    from linear fits for each of the three classes
    of spheres. (c) Experimental measurements of particles' 
    axial scale factors,
    $m_p(a_p)$,
    as a function of
    radius, $a_p$, for spheres made
    of polystyrene, silica and TPM.
    Values from (a) are plotted as crosses.}
    \label{fig:zdependence}
\end{figure*}

The representative data plotted 
in Fig.~\ref{fig:hothvm}(c) illustrate
the challenge posed by axial displacements
in holographic optical traps.
The two traces show the axial positions of
a polystyrene sphere (PS, ThermoFisher Scientific, catalog number 4202A) and a silica
sphere (SiO$_2$, Bangs Laboratories, catalog number SS05N) localized in water
by an optical tweezer
and translated straight upward along
the optical axis in a sample cell composed
of two parallel glass surfaces separated by
\SI{15}{\um}.
The two spheres were selected from
two populations co-dispersed
in the same sample at a concentration of
\SI{e5}{particles\per\milli\liter}.
The polystyrene sphere has a measured
radius of $a_p = \SI{0.989(1)}{\um}$
and refractive index $n_p = \num{1.612(1)}$.
The silica sphere
has a radius of $a_p = \SI{1.184(31)}{\um}$
and refractive index $n_p = \num{1.396(3)}$.
The two particles are trapped and translated
one at a time by the
same optical tweezer
located at a fixed in-plane position within the
sample cell.
The trapped sphere is raised in discrete
steps of \SI{750}{\nm}.
Its characteristics and mean position 
at each step are obtained
by analyzing \num{50} holograms recorded
at \SI{24}{frames\per\second} to 
average over thermal fluctuations.
The holograms in Fig.~\ref{fig:hothvm}(b)
show individual snapshots from each of three
stages in this process.

The trapped sphere eventually collides with 
the upper glass wall of the sample cell
and so stops rising, even as the trap
moves further upward.
The height of the plateau augmented by the
holographically measured radius of the sphere provides an estimate for the
axial position of the wall, 
$z_\text{wall} = \max(z_p(z_j)) + a_p$,
as indicated in Fig.~\ref{fig:hothvm}(c).
Although the two spheres' plateau heights
differ by \SI{240(40)}{\nm}, the associated estimates
for $z_\text{wall}$ differ by just \SI{40}{\nm},
which is comparable
to the uncertainty in the radius of the
silica particle.
Comparably good agreement is
obtained consistently with
different pairs of particles
and usefully validates
the precision and accuracy of Lorenz-Mie
microscopy for measuring particles'
axial positions and radii.

Using $z_\text{wall}$ as a fiducial point,
the substantial
difference of \SI{1.2}{\um} in the spheres' 
axial positions just before they reach
the wall at $z_j = \SI{14.5}{\um}$ cannot
be ascribed to measurement error.
Instead, this discrepancy shows that the two spheres
rise through the cell with significantly
different values of the axial scale factor,
\begin{equation}
\label{eq:scalefactor}
m_p \equiv \left<\frac{d z_p}{d z_j}\right>.
\end{equation}
The polystyrene sphere rises faster
than the trap that is translating it, with
$m_p = \SI{1.068(7)}{\um\per\um}$.
The silica sphere
sits consistently lower in the
trap
and moves upward in significantly
smaller steps, with
$m_p = \SI{0.967(7)}{\um\per\um}$.

Comparable discrepancies in $m_p$
are observed reproducibly, 
not just from particle to particle at the same position
in the same sample cell, but also from position to
position and even in different sample cells.
Figure~\ref{fig:zdependence}(a) presents axial
translation data from three types of colloidal
spheres measured in multiple sample
cells.
These samples consist of
\SI{1.0}{\um}-radius polystyrene (PS) spheres,
\SI{0.8}{\um}-radius 3-(trimethoxysilyl)propyl methacrylate (TPM) spheres \cite{hannel18} and
\SI{0.8}{\um}-radius silica (SiO$_2$) spheres.
Each type of sphere translates
with a distinctive axial scale factor.
The solid lines in Fig.~\ref{fig:zdependence}(a)
represent linear least squares fits
whose slopes yield $m_p$.
Their residuals, plotted in Fig.~\ref{fig:zdependence}(b),
display no significant
trends.

The axial scale factors
from Fig.~\ref{fig:zdependence}(a)
are plotted as crosses
in  Fig.~\ref{fig:zdependence}(c),
where they 
are compared with corresponding results
from other populations of spheres
made of the same three materials.
Micrometer-scale spheres
made of polystyrene  and TPM
both rise faster than
their traps ($m_p > 1$), 
with axial scale factors
that depend only weakly on
size.
Silica spheres have axial scale factors
that decrease significantly with
increasing particle radius,
smaller spheres rising
faster than their traps
and larger spheres rising
more slowly.

\section{Modeling axial displacements}
\label{sec:axial}

The observed deviations of $m_p$ from
unity cannot arise from a simple scaling error
because different types of spheres consistently
deviate by different characteristic amounts.
Nor can these deviations be attributed to aberrations in 
the trapping beam because both measurements 
reported in Fig.~\ref{fig:hothvm}(c) were
performed at the same position in the same
sample cell, without any intervening
mechanical adjustments
that could have changed the shape of the trap.

To explain these observations,
we model the intensity profile of
an optical tweezer
as a Gaussian beam brought to a
focus at $\vec{r}_j$ with
intensity profile, $\abs{\vec{E}(\vec{r} - \vec{r}_j)}^2$, given
by
\begin{equation}
    \label{eq:intensity}
    \abs{\vec{E}(\vec{r})}^2
    =
    \abs{E_j}^2
    \frac{z_R^2}{z^2 + z_R^2} \,
    \exp\left( 
    -2 \frac{r^2}{w_0^2} \frac{z^2}{z^2 + z_R^2}
    \right),
\end{equation}
where $w_0 \approx \lambda_0 / (n_m \, \text{NA})$ is the
radius of the beam waist.
The axial extent of the
focal spot depends on how
tightly the beam is focused
and is described by
the Rayleigh range, $z_R$.
A small particle with complex dipole polarizability
$\alpha_e = \alpha_e' + i \alpha_e''$ 
displaced from the focal point by
$\vec{r} = (r, \theta, z)$
experiences a dipole-order 
force \cite{yevick17},
\begin{equation}
    \label{eq:F_e}
    \vec{F}_e(\vec{r}) 
    = 
    - \frac{1}{2} \, k \alpha_e' \,
    \abs{E_j}^2 \,
    \left[\frac{r}{z_R} \, \hat{r} +
    \frac{z - z_0}{k z_R^2} \, \hat{z}
    \right],
\end{equation}
that draws it toward the axis at a distance,
\begin{equation}
    \label{eq:z0}
    z_0 = \frac{\alpha_e''}{\alpha_e'} z_R (kz_R - 1) ,
\end{equation}
downstream of the focal point.
This displacement is independent
of the trap's intensity, $\abs{E_j}^2$, and reflects
a balance between the 
dipole-order restoring force
arising from
intensity gradients
and radiation pressure directed
by phase gradients.

\begin{figure*}
    \centering
    \includegraphics[width=\textwidth]{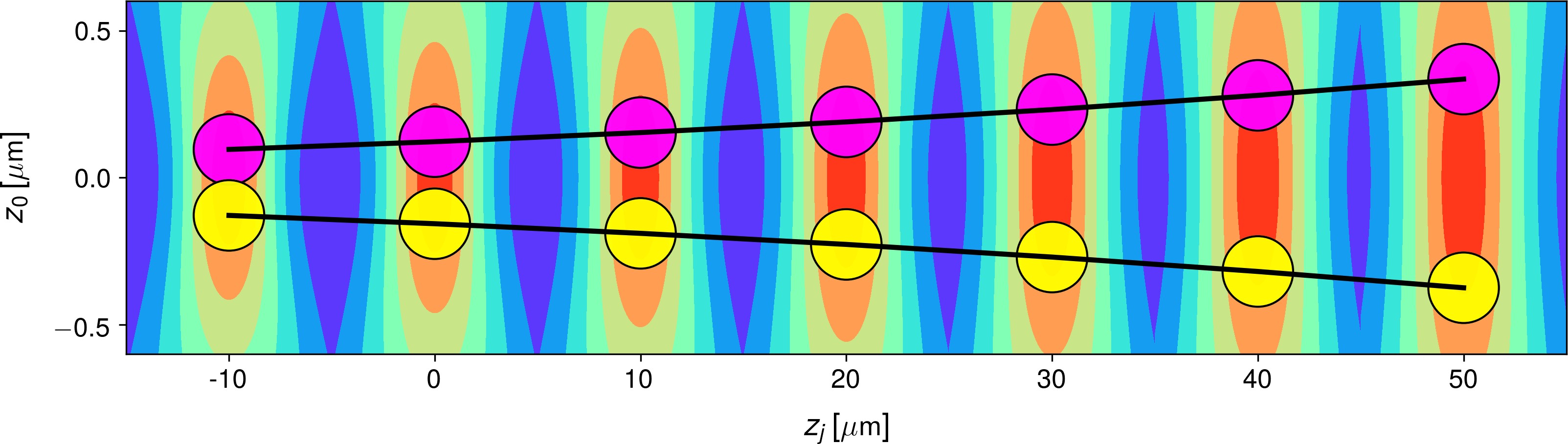}
    \caption{Particle position, $z_0$, within an optical trap as a function of the trap's axial displacement, $z_j$ for \SI{120}{\nm}-diameter polystyrene (PS, magenta, upper) and
    silica (SiO$_2$, yellow, lower) spheres dispersed in water and trapped in
    a diffration-limited optical tweezer
    at $\lambda_0 = \SI{1064}{\nm}$.
    Colored contours reveal how
    the trap's intensity profile,
    $\abs{\vec{E}(\vec{r})}^2$, broadens and elongates with increasing $z_j$.
    The polystyrene sphere rises in the
    trap, while the silica sphere sinks.}
    \label{fig:forcebalance}
\end{figure*}

Moving the trap along the
axis by $z_j$
changes $z_0$
by changing the shape
of the beam's focus.
Specifically, the Rayleigh range
depends on $z_j$ as
\cite{self1983focusing}
\begin{equation}
    \label{eq:rayleighrange}
    z_R(z_j) = z_R(0) \left(1 + \frac{z_j}{f}\right)^2,
\end{equation}
where $z_R(0) \approx 8 / (k \text{NA}^2)$
sets the scale for a Gaussian beam focused by a lens of numerical aperture NA.
The derivation of Eq.~\eqref{eq:rayleighrange} requires the Rayleigh range
of the beams diffracted by the SLM to be much larger than the focal length
of the objective lens, a condition that generally is 
met in holographic trapping systems.

In the absence of other forces, a trapped particle 
comes to mechanical equilibrium at 
\begin{equation}
\label{eq:zp}
z_p(z_j) = z_j + z_0(z_j) .
\end{equation}
Particles therefore tend to be displaced along the optical axis
by more than the displacement of the optical trap, with
the extra displacement depending on the particle's size and
refractive index.
The upper trace in Fig.~\ref{fig:forcebalance}
shows $z_0(z_j)$ for
a \SI{120}{\nm}-diameter polystyrene sphere
in an optical trap modeled by Eq.~\eqref{eq:intensity}
with parameters for our instrument.
Contours of the intensity distribution show
that the focal spot expands along the axial
direction as $z_j$ increases, thereby increasing
the particle's displacement from the focal point.

Displacing a trap along the axial direction
decreases its stiffness through the dependence 
of $\vec{F}_e(\vec{r})$ on $z_R$, as shown approximately in
Eq.~\eqref{eq:F_e}.
The scale of a trapped particle's 
thermal fluctuations about $z_p(z_j)$
therefore should increase as $z_j$ increases.
This trend is reflected in the increasing
standard deviation of $\Delta z_p(z_j)$ plotted in
Fig.~\ref{fig:zdependence}(b).

For displacements smaller than
the focal length of the
objective lens, $z_j < f$,
the dependence of $z_p$ on $z_j$ is 
roughly linear and is characterized by the
axial scale factor
\begin{equation}
    \label{eq:mppredicted}
    m_p
    \approx
    1 + 2 \, \frac{\alpha_e''}{\alpha_e'} \,
    \frac{z_R(0)}{f}
    [2 k z_R(0) - 1].
\end{equation}
For our instrument,
$m_p \approx 1 + 0.037 \frac{\alpha_e''}{\alpha_e'}$.
Observing the nonlinear scaling
predicted by Eqs.~\eqref{eq:z0}
through \eqref{eq:zp} is not
feasible because particles
escape their traps
when $z_0(z_j)$ becomes large.
Indeed, material-dependent displacements
set an upper limit on the axial range through
which colloidal 
particles can be translated with holographic
optical tweezers.

For the particular case of a small dielectric sphere of 
radius $a_p$ and refractive index $n_p$
immersed in a medium of refractive index $n_m$, 
the electric dipole polarizability is \cite{draine93}
\begin{equation}
    \label{eq:alpha}
    \alpha_e
    =
    \frac{\alpha_e^{(0)}}{1 - \frac{i}{6 \pi \epsilon_0 n_m^2} k^3 \alpha_e^{(0)}},
\end{equation}
where the Clausius-Mossotti polarizability is
\begin{equation}
    \label{eq:alpha0}
    \alpha_e^{(0)} = 4 \pi \epsilon_0 n_m^2 \, a_p^3 \,
    \frac{n_p^2 - n_m^2}{n_p^2 + 2 n_m^2} .
\end{equation}
Given this, a sphere's axial scale factor depends
on its radius, $a_p$, and refractive index, $n_p$,
through the ratio
\begin{equation}
    \label{eq:z0scale}
    \frac{\alpha_e''}{\alpha_e'}
    \approx
    \frac{2}{3} (k a_p)^3 \, 
    \frac{n_p^2 - n_m^2}{n_p^2 + 2 n_m^2}.
\end{equation}
Small, weakly scattering particles therefore
tend to be localized near the focal points
of their traps.
Variability in axial
placement becomes more of
an issue for larger particles
with larger index mismatches.

Equation~\eqref{eq:z0scale}
is valid for particles that
are substantially smaller
than the wavelength of
light $k a_p < 1$.
Holographic tracking and
characterization, however,
requires particles that
are larger than the wavelength
of light.
The data for polystyrene and TPM
in Fig.~\ref{fig:zdependence}(c)
therefore all have $k a_p > 1$ and
show a comparatively weak
dependence on particle size,
presumably because the particles
are larger than the radius of the beam waist,
$w_0 \approx \SI{550}{\nm}$.

The data for silica spheres in Fig.~\ref{fig:zdependence}(c)
show a downward trend in $m_p(a_p)$ that runs counter
to the prediction of Eq.~\eqref{eq:mppredicted} and can be
explained by the influence of gravity.
The weight of a sphere of density $\rho_p$
dispersed in a fluid of density $\rho_m$
shifts the point of
mechanical equilibrium by
a distance
\begin{equation}
    \label{eq:deltaz0}
    \Delta z_0(z_j) 
    = 
- \frac{8}{3} \pi a_p^3 \, (\rho_p - \rho_m) \, g \frac{z_R^2(z_j)}{\alpha_e'\abs{E_j}^2},
\end{equation}
where $g$ is
the acceleration due to gravity.
This offset depends on axial displacement
through the dependence of $z_R$ on $z_j$,
and thus affects the axial scale factor, $m_p$.
The lower trace in Fig.~\ref{fig:forcebalance}
shows how the axial
displacement of a \SI{120}{\nm}-diameter silica sphere
depends on trap position given the combination of
optical and gravitational forces.
Unlike the neutrally-buoyant polystyrene sphere
($\rho_p = \SI{1.05}{\gram\per\cubic\cm}$)
the dense silica sphere 
($\rho_p = \SI{2}{\gram\per\cubic\cm}$)
lags behind the focal point
as the trap moves upward.
TPM has an intermediate refractive index
($n_p = \num{1.5}$) and an intermediate
density ($\rho_p = \SI{1.23}{\gram\per\cubic\cm}$)
and so is predicted to have an
intermediate displacement.

Equation~\eqref{eq:deltaz0} is valid for particles
whose density mismatch is small enough that 
$z_0 + \Delta z_0 < z_R$.
Within this range, Eqs.~\eqref{eq:alpha}, \eqref{eq:alpha0} and
\eqref{eq:deltaz0} suggest that $\Delta z_0$ depends only
weakly on $a_p$.
In fact, the data for silica
spheres in Fig.~\ref{fig:zdependence}(c)
shows that gravity causes larger
spheres to sit lower in their
traps and to rise more slowly.
This is consistent with the observation
of weaker-than-predicted size scaling
for comparably sized polystyrene and TPM spheres.
Although the dipole-order theory accounts
for qualitative features of the
observed displacements, quantitative 
predictions for larger spheres presumably
require a higher-order treatment.
Limitations of the analytically tractable dipole-order
theory highlight the value of holographic microscopy
for providing \emph{in situ} experimental feedback,
particularly for particles that are
larger than the wavelength of light.

\section{Discussion}
\label{sec:discussion}

We have used Lorenz-Mie microscopy to 
demonstrate that holographically trapped colloidal
spheres are displaced within their traps by amounts
that depend substantially on the traps' axial positions.
These displacements, often amounting to more than the
wavelength of light, are explained by changes in the
traps' Rayleigh ranges as they move along the
optical axis. How this affects the position of a
particle within a trap depends on the particle's
size and refractive index, and also can be influenced
by external forces such as gravity.
Whereas in-plane displacements can be measured
and calibrated with conventional microscopy
and standard techniques of image analysis \cite{crocker96},
axial displacements have been much more challenging
to measure and so have been largely overlooked.
The present study demonstrates that ignoring
material-dependent axial displacements 
can lead to large errors in particle placement,
not only for heterogeneous assemblies, but also
for nominally identical spheres arranged in
three-dimensional patterns.
The comparative difficulty of predicting and
correcting these offsets \emph{a priori}
creates a need for the real-time feedback that can be provided
by quantitative holographic video microscopy.

\section*{Funding}

This work was supported by the MRSEC program of the
National Science Foundation through Award Number
DMR-1420073.
The instrument used for this work  was constructed with support of the MRI program of the NSF under Award Number DMR-0922680.
We gratefully acknowledge the support of the nVidia Corporation through the donation of the Titan Xp GPU used for this research.

\bibliography{zrange}

\end{document}